\documentclass[]{IEEEtran}
\IEEEoverridecommandlockouts
\input epsf
\usepackage{cite}
\usepackage{graphicx}
\usepackage{url}
\usepackage{xcolor}
\usepackage{upgreek}

\usepackage{amsmath}
\usepackage[colorlinks=true,linkcolor=blue,urlcolor=blue,citecolor=blue]{hyperref}
\usepackage[colorinlistoftodos]{todonotes}

\begin{document}


\IEEEpubid{\makebox[\columnwidth]{979-8-3315-4683-0/26/\$31.00~\copyright2026 IEEE \hfill} \hspace{\columnsep}\makebox[\columnwidth]{ }}

\title{\LARGE Design and Initial Test of the Weighing Unit for Tsinghua Tabletop Kibble Balance}


\author{Weibo Liu$^1$,
Nanjia Li$^1$, Lushuai Qian$^2$, Wei Zhao$^1$, Lisha Peng$^1$, Songling Huang$^1$, Shisong Li$^\dagger$\\
1. Department of Electrical Engineering, Tsinghua University, Beijing 100084, China\\
2. China Jiliang University, Hangzhou 310018, China\\
$^\dagger$Email: shisongli@tsinghua.edu.cn} 


\maketitle

\IEEEpubidadjcol

\begin{abstract}
This paper presents a customized weighing unit developed for the Tsinghua tabletop Kibble balance. The system is based on a flexure hinge mechanism sourced from a commercial weighing cell, with a major modification to the feedback control loop. The redesigned loop incorporates a capacitive displacement sensor for high-resolution position detection and a novel PID control strategy that ensures both fast dynamic response and high static stability. Initial characterization results demonstrate a repeatability better than 0.1\,mg in air for 1\,kg mass exchanges, validating the system's potential for high-accuracy mass metrology in Kibble balances. 
\end{abstract}


\pagenumbering{gobble}

\vspace{-0.5cm}
\section{Introduction}

The Kibble balance~\cite{Kibble1976} has become a fundamental method for realizing the kilogram following the 2019 redefinition of the International System of Units (SI). Compact tabletop Kibble balances have emerged as a new trend in recent years~\cite{NISTTabletop2024,NPLCPEM2024,PTB2025}. The Tsinghua tabletop Kibble balance (THUKB)~\cite{THUdesign2022,THU2025updates} is designed as a compact, robust, and cost-effective instrument, targeting a relative uncertainty of $10^{-8}$ at 1\,kg—a level that would enable mass calibration capabilities comparable to those of National Metrology Institutes (NMIs).

The weighing unit is a critical subsystem, as its performance directly governs the achievable resolution and accuracy of the entire balance. Although commercial electromagnetic force compensation (EMC) weighing cells offer well-established metrology, their fixed control parameters and the high cost of custom modifications render them ill-suited for the specific demands of Kibble balance operation. To address these limitations, this paper presents a dedicated weighing unit developed by modifying the feedback loop of a commercial EMC cell. We describe the mechanical and control system architecture and report preliminary performance tests that validate the design's potential for high-accuracy mass metrology in a compact form factor.

\IEEEpubidadjcol

\section{Design of the weighing unit for THUKB}

\begin{figure}[tp!]
    \centering    \includegraphics[width=0.44\textwidth]{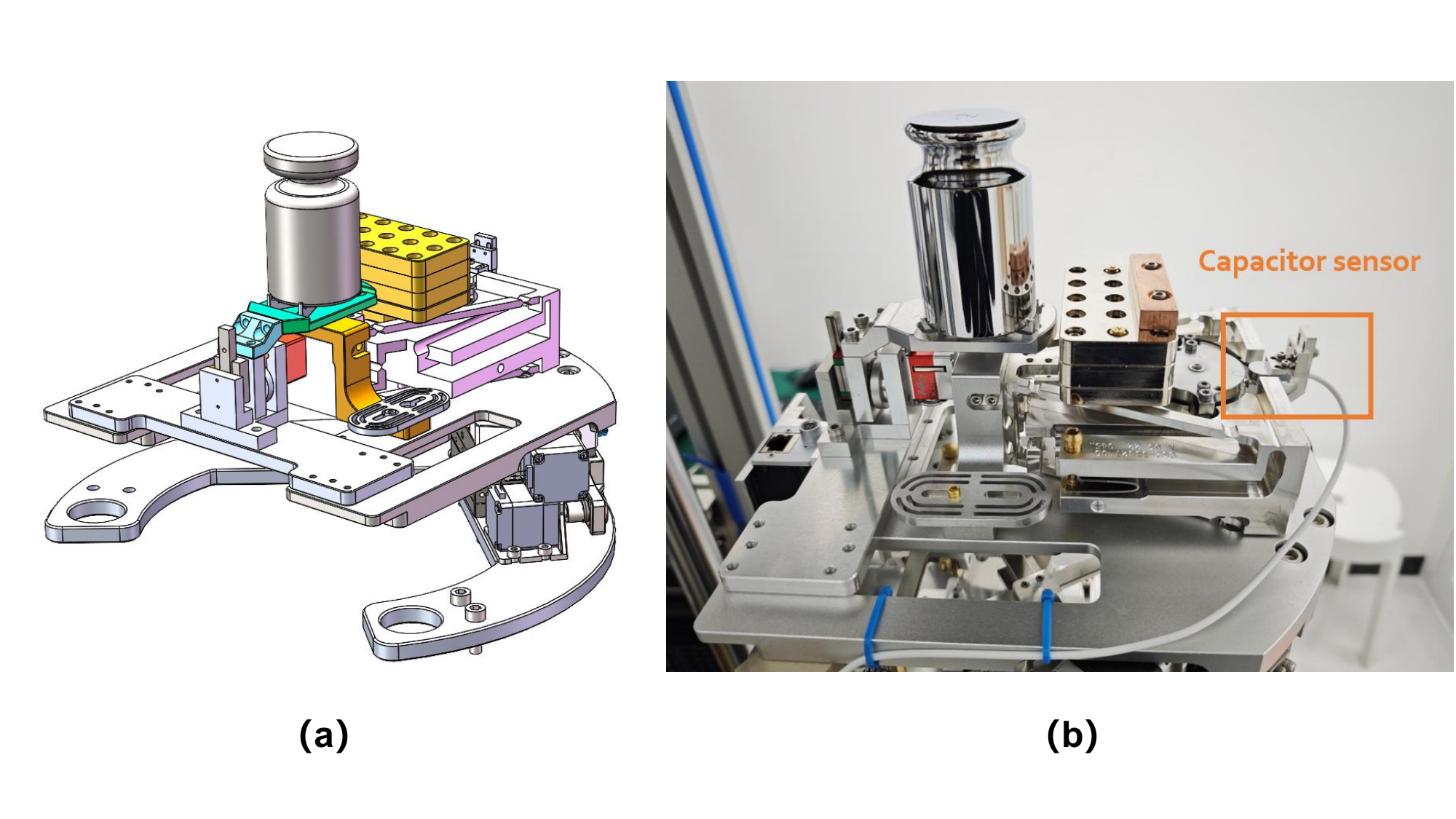}
    \caption{Weighing Unit
for THUKB. (a) CAD model. (b) Initial setup.}
    \label{fig:weighingDesign}
\end{figure}

The weighing unit for THUKB is adapted from the Sartorius WZA215-LC, which originally provides a capacity of 210\,g, a readability of 10\,$\upmu$g, and a reproducibility of better than $\pm$20\,$\upmu$g. Key mechanical modifications are shown in Fig. \ref{fig:weighingDesign}, including adding a 1\,kg mass pan, a flexure-based coil suspension system, a mass exchanger for automated 1\,kg mass-on/off operations, an extended mass pan for corner error adjustment, and a counter-mass assembly. Under a total dead weight of approximately 3\,kg, the weighing unit maintains structural and functional integrity. To enhance displacement sensing resolution, the original optical sensor of the commercial weighing cell was replaced with a high-resolution capacitive sensor (range $400\,\upmu$m, resolution 0.35\,nm), shown in Fig. \ref{fig:weighingDesign}(b), providing superior feedback for position control. A detailed comparison of the sensors can be found in \cite{THU2025updates}.

\begin{figure}[tp!]
    \centering    \includegraphics[width=0.42\textwidth]{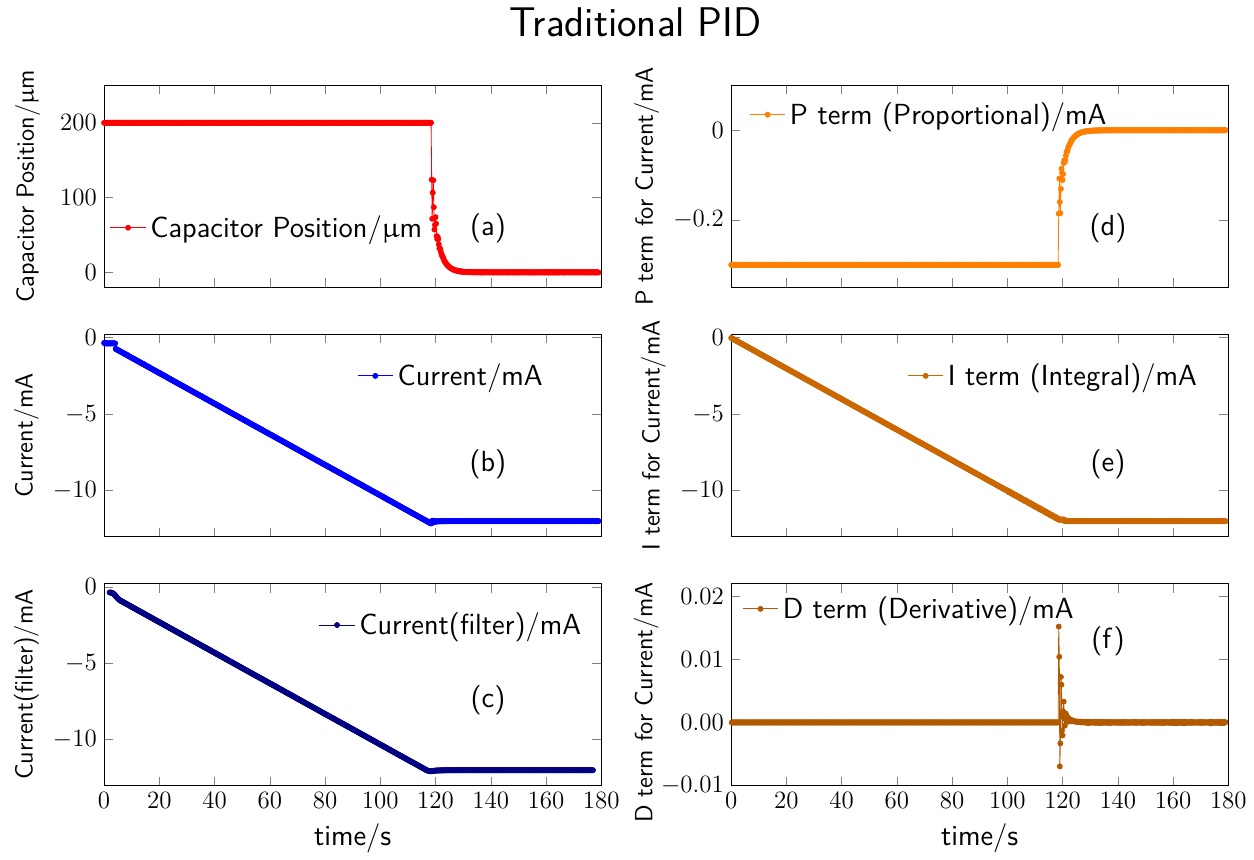}
    \caption{Using traditional PID strategy, the measured current ramping time when weighing 1\,kg is 120\,s. In (c), a moving average filter is applied to reduce the interference of noise in the measurements.}
    \label{fig:TraditionalPID}
\end{figure}

In the weighing phase, a PID controller maintains the weighing mechanism at a null position. The displacement error $e(t) = z(t) - z_{\mathrm{ref}}$ is measured by the capacitive sensor, and the controller output $i(t)$ drives the coil to compensate for the applied mass load~\cite{THU2024currentsource}. The current ramping time for the traditional PID, which aims to maintain good static stability, is exceptionally long at 120\,s, as shown in Fig.~\ref{fig:TraditionalPID}, due to the integration term $K_i\int e(t)dt$ ($K_i$ is set small for improving the static performance).

\begin{figure}[htbp!]
    \centering    \includegraphics[width=0.42\textwidth]{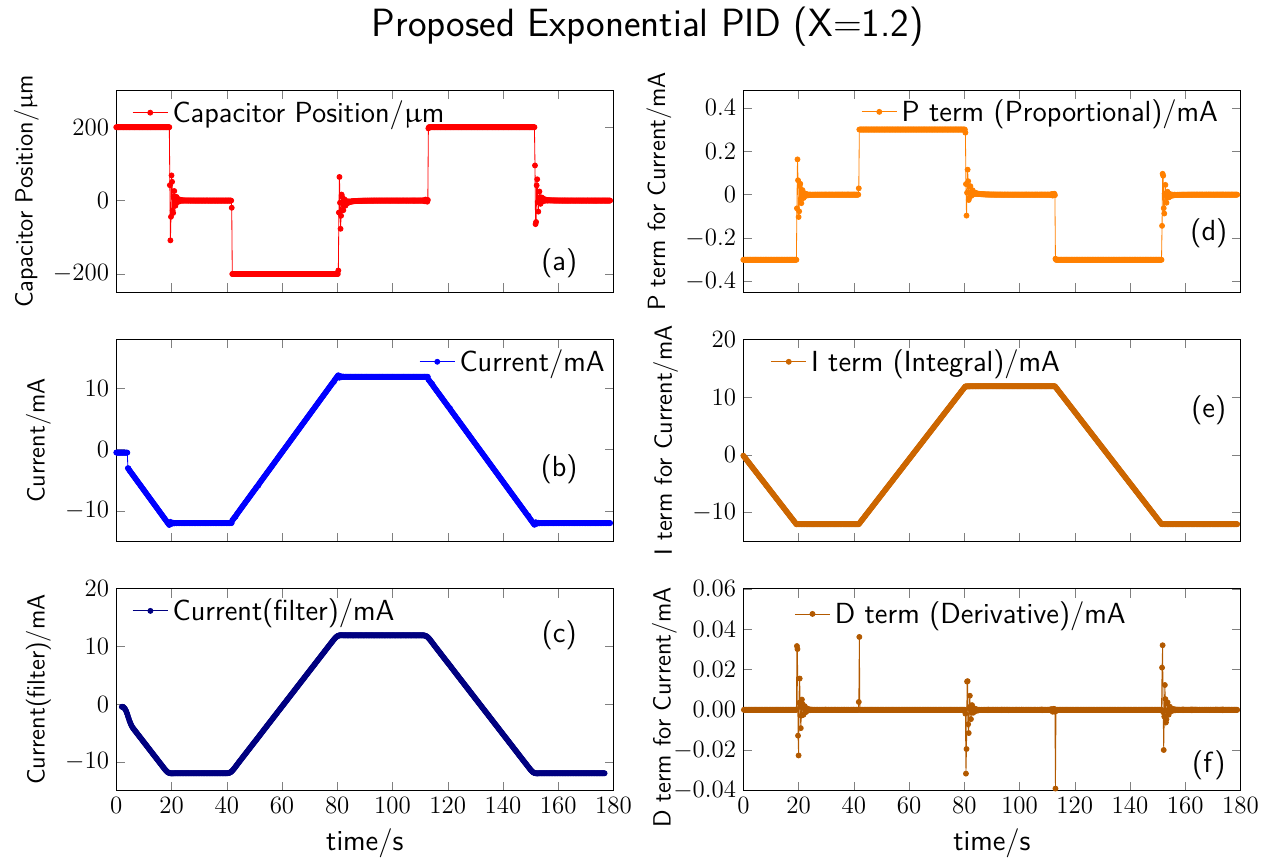}
    \caption{The proposed exponential PID measurement yielded an current ramping time of 40\,s when weighing 1\,kg.}
    \label{fig:ProposedPID}
\end{figure}

To achieve dynamic rapid response and steady-state low-noise characteristics in a smooth transition process, a novel exponential integral PID control strategy is proposed as 
\begin{equation}
i(t) = K_p e(t) + K_i\int X^{\lvert e(t) \rvert} e(t)dt + K_d \frac{d e(t)}{dt}.
\label{eq:PID}
\end{equation}
where $X$ is a base greater than 1, $K_p$, $K_i$, and $K_d$ respectively the proportional, integral, and derivative coefficients. When the error $e(t)$ magnitude is substantial, the integral component undergoes exponential amplification, accelerating the current ramping to 40\,s with $X=1.2$ as in Fig.~\ref{fig:ProposedPID}. In steady-state operation, the controller functions identically to conventional PID systems.
\vspace{-0.3cm}

\section{Initial Weighing experiment}
\begin{figure}[htbp!]
    \centering    \includegraphics[width=0.28\textwidth]{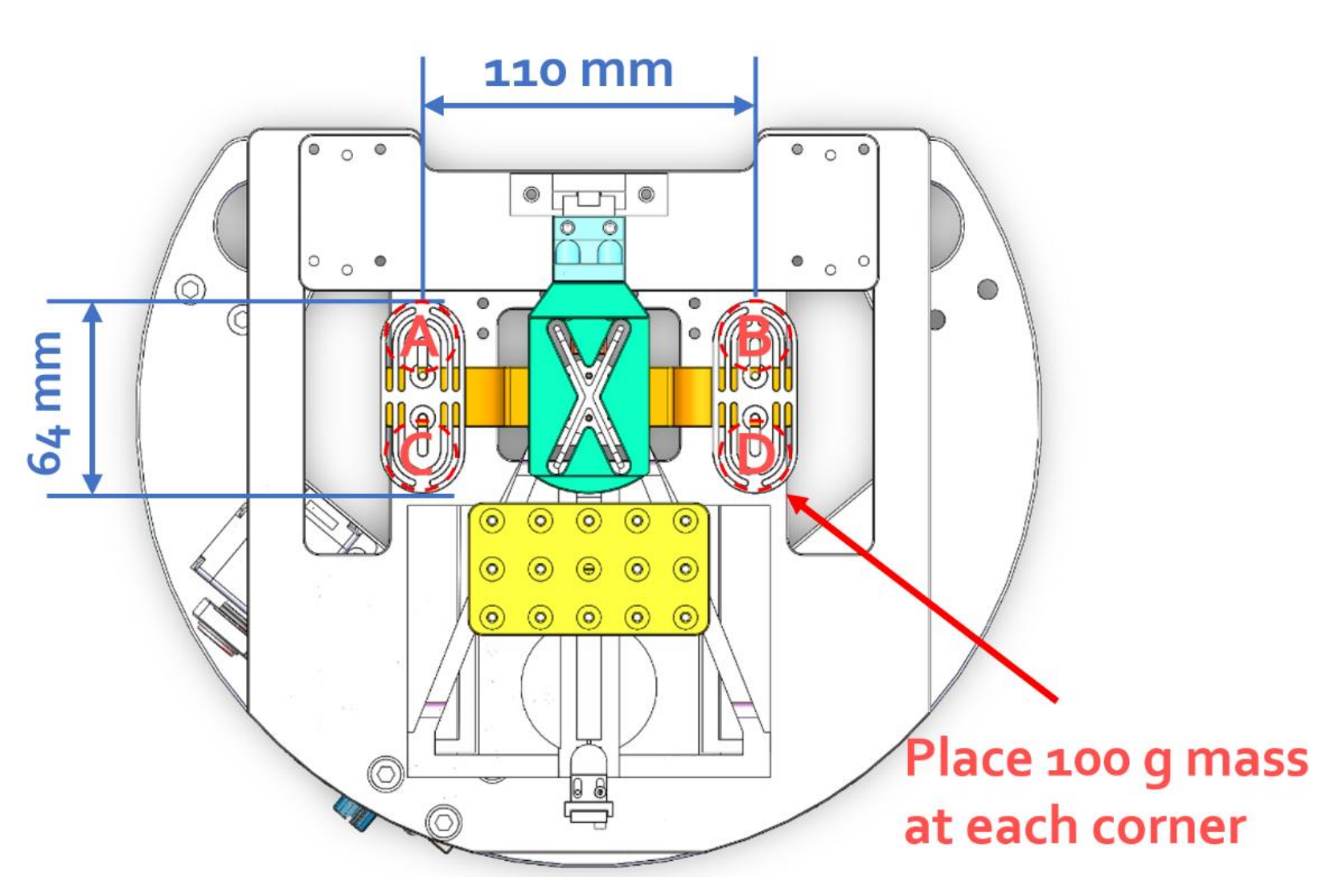}
    \caption{Mechanical design for the corner error adjustment.}    \label{fig:FigCPEM2026_Setup}
\end{figure}

Corner error adjustment constitutes a critical preliminary procedure in the weighing experiment. Without proper correction, minute displacements of the mass pan during 1\,kg mass-on/off operations can introduce significant deviations in the mass readout. To mitigate this effect, an extended mass pan was designed as in Fig.~\ref{fig:FigCPEM2026_Setup}, enabling the placement of a 100\,g test mass at each of the four corners sequentially. The corresponding mass value, derived from current feedback, is recorded for each position. After the initial adjustment, the measured mass values at all corners agree within 1\,mg consistency.

\begin{figure}[htbp!]
    \centering
    \includegraphics[width=0.32\textwidth]{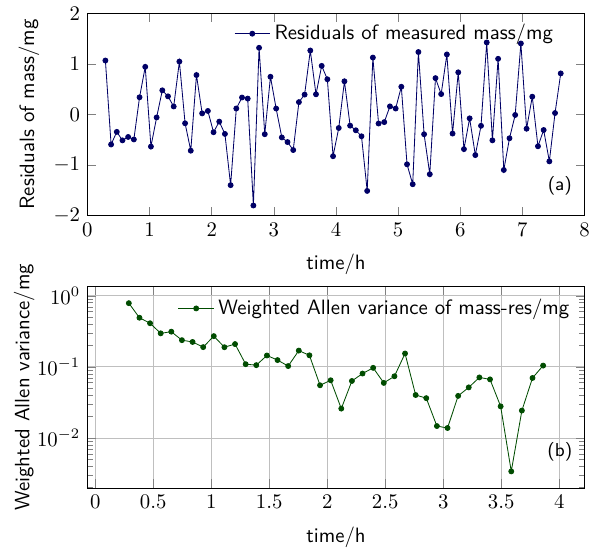}
    \caption{The initial 8-hour repeatability result of 1\,kg mass-on/off in air.}
    \label{fig:FigCPEM2026_repetability}
\end{figure}

An initial 8-hour repeatability test of 1\,kg mass-on/off measurements was conducted under ambient atmospheric conditions with the vacuum chamber closed. The mass was exchanged every 80\,seconds (40\,seconds per reading state), yielding 11 complete measurement cycles per hour. Data were processed using an ABA sequence (mass-off, mass-on, mass-off) to mitigate $Bl$ drift, and the test results presented in Fig.~\ref{fig:FigCPEM2026_repetability} show the residual mass variation over the 8-hour period was approximately 2\,mg peak-peak value. The weighted Allan deviation of the data reached a repeatability better than 0.1\,mg after an averaging time of 2 hours.

\section{Conclusion}
A custom weighing unit for the Tsinghua tabletop Kibble balance has been developed and experimentally validated. By integrating a high-resolution capacitive displacement sensor together with the proposed exponential PID control strategy, the system achieves a significant enhancement in weighing performance. Initial characterization results demonstrate a repeatability of better than 0.1\,mg for 1\,kg mass exchanges in air, confirming the effectiveness of the modified feedback loop. Future work will focus on further performance optimization and full characterization of the weighing unit.

\section*{Acknowledgment}
This work was supported by the National Natural Science Foundation of China under Grant 52377011.


\end{document}